# Facet Dependent Topological Phase Transition in Bi$_4$Br$_4$

Jingyuan Zhong[1], Ming Yang[1], Fei Ye[1], Chen Liu[2], Jiaou Wang[2], Weichang Hao[1], Jincheng Zhuang[1], Yi Du[1]

[1] School of Physics, Beihang University, Haidian District, Beijing 100191, China

[2] Beijing Synchrotron Radiation Facility, Institute of High Energy Physics, Chinese Academy of Sciences, Beijing 100049, P. R. China

Jingyuan Zhong and Ming Yang contributed equally to this work. Correspondence and requests for materials should be addressed to J.C.Z. (email: jincheng@buaa.edu.cn) or to W.C.H. (email: whao@buaa.edu.cn)　or to Y.D. (email: 10154@buaa.edu.cn)

**Abstract:** The realization of the coexistence of various topologically nontrivial surface states in one material is expected to lay a foundation for new electric applications with selective robust spin current. Here we apply the magnetoconductivity characteristic and angle-resolved photoemission spectroscopy (ARPES) to visualize the surface-selected electronic features evolution of quasi-one-dimensional material Bi$_4$Br$_4$. The transport measurements indicate the quantum interference correction to conductivity possesses symbolic spin rotational characteristic correlated to the value of Berry phase with the effects of weak localization and weak antilocalization for (001) and (100) surfaces, respectively. The ARPES spectra provide the experimental evidence for quasi-one-dimensional massless Dirac surface state at the side (100) surface and anisotropic massive Dirac surface state at the top (001) surface, respectively, which is highly coincide with the angle-dependent scaling behavior of magnetoconductivity. Our results reveal the facet dependent topological phases in quasi-one-dimensional Bi$_4$Br$_4$, stimulating the further investigations of this dual topology classes and the applications of the feasible technologies of topological spintronics.

Searching for topological quantum materials (TQMs) has become a hot topic in the region of condensed matter physics and materials science due to their underlying mechanisms in physical basis and potential applications of surface states protected by various symmetries, such as time-reversal symmetry (TRS) and lattice symmetries, serving as robust channels against the external perturbations in low consumption electronics and high performance spintronics. Since these TQMs possess various exotic phenomena, many works have been dedicated to realize the topological phase transition by modulating these topological properties to understand the correlations among different topologies and to promote their practical applications. The widely used mean to achieve topological phase transition is chemical substitution, where an odd or even band inversion could be evoked by simultaneously varying the spin-orbit coupling (SOC) strength and lattice constant[1-8]. Nevertheless, the different topological phases reside in a series of samples rather than in one sample, requiring the additional integration of different samples in the device applications. Recently, the strain effect is applied to enable the reversible control of the topological phase transition by tuning the lattice constant[9,10]. The necessity of precise control of the strain strength, however, provides the obstacle to hinder their potential applications. Therefore, exploration of a material with natural coexistence of different topological phases originating from the bulk band inversions is highly demanded for technological implications.

The bismuth halide, $Bi_4X_4$ ($X$ = I or Br), possess the unique structure with one-dimensional (1D) infinite molecular chain as the structural building block to stack through van der Waals force for both of their two-dimensional (2D) and three-dimensional (3D) allotropes. Due to the high atomic number of Bi element, the large SOC strength in this system is expected to give rise to the band inversion. In fact, a large-gap quantum spin Hall state in $Bi_4Br_4$ resulted from the SOC of Bi $2p$ orbits has been theoretically predicted by first-principle calculations and then experimentally confirmed by scanning tunnelling spectroscopy[11-16]. This system holds different out-of-plane stacking orders, bringing in the different lattice symmetries on top of TRS and leading to a variety of topological phases, including weak topological insulator (WTI) and high-order topological insulator (HOTI)[17-20]. Furthermore, the 3D $α'$-$Bi_4Br_4$ is predicted to be either a topological crystalline insulator or HOTI form the view of both TRS and lattice symmetries[21]. All these results indicate that the bulk $Bi_4Br_4$ could be a promising candidate possessing various topologies.

In this work, we demonstrate the quantum interference correction to conductivity reflected by coexistence of weak localization (WL) and weak antilocalization (WAL) correlated to top (001) surface and side (100) surface, respectively, in single crystal $Bi_4Br_4$ by the angle-dependent magnetic conductivity measurements. Both of WL and WAL originate from the electron coherent backscattering between two time-reversal states with different Berry phases ($\varphi$) restricted by spin rotational characteristic. The angle-resolved photoemission spectroscopy reveals the massive Dirac fermions ($\varphi \neq \pi$) at the (001) surface and massless Dirac cone ($\varphi = \pi$) at (100) surface, coinciding with magnetic conductivity results and implying facet dependent topological phase transition in $\alpha'$-$Bi_4Br_4$.

## Results

**Coexistence of WL and WAL phenomena.** Fig. 1(a) displays the schematic diagram of atomic structure of $\alpha'$-$Bi_4Br_4$, where the structural building block is the 1D infinite chain along $b$ axis with four Bi atoms in width and terminated by Br atoms. Both of the intralayer interaction along $a$ axis and interlayer interaction along $c$ axis are van der Waals force, resulting in needle-like shape of single crystal $Bi_4Br_4$ as shown Supplementary Note 1. Consequently, the conductivity of single crystal $Bi_4Br_4$ is made up of three parts: bulk conductivity, (001) surface conductivity, and (100) surface conductivity, which is distinct with the conductivity of 2D layered materials contributed by the bulk conductivity and only one cleavable surface[22-30]. Supplementary Note 2 shows the total magnetoconductivity (MC), defined as MC = 1/R(B) – 1/R(B = 0), where R(B) is the magneto-resistivity, of single crystal $Bi_4Br_4$ measured at 2 K with the angle $\alpha$ between the applied magnetic field ($B$) and the current, as shown in the inset of Fig. 1(b). The direction of magnetic field is restricted in the $bc$ plane of $Bi_4Br_4$ during the sample rotation. All the MC curves display the similar trend with only subtle variation. In order to figure out the role of surface states in the MC, we subtract the contribution of bulk states applying the formula: $\Delta\sigma(\alpha, B) = \Delta\sigma(\alpha, B) - \Delta\sigma(90°, B)$, by assuming the isotropic MC from bulk states[23,29]. Fig. 2(b) displays the $\Delta\sigma$ as a function of normal $B$ component ($B\cos\alpha$) at different $\alpha$ with the temperature range from 2 K to 20 K. At low temperature, all the angle-scaled curves coincide with each other at low magnetic field and show a negative cusp, which is interpreted as the presence of WAL in topological insulators[22-24,31-33]. The spin-momentum locked Dirac fermions resulted from the gapless Dirac cone in topological insulators accumulate a π Berry phase after going across two time-reversed self-intersecting trajectories, leading to the transport quantum interference from constructive to destructive[24,34,35]. This destructive quantum interference enhances the conductivity by prohibiting the return probability of the Dirac fermions, which is suppressed by the perpendicular breaking the TRS and π Berry phase, leading to a negative MC with the cusp in Fig. 2(b). With the increment of magnetic field, an upward tendency has been identified, which is the characteristic of WL effect with the constructive quantum interference between two scattering loops[24].

In order to make a quantitative analysis of the WAL effect on the MC curves at low magnetic field, the Hikami-Larkin-Nagaoka (HLN) theory is applied to describe MC variation by the formula[31]:

$$\Delta\sigma = \varepsilon \left(\frac{e^2}{\pi h}\right)[\psi\left(\frac{1}{2} + \frac{h}{8\pi e l_\phi^2}\right) - \ln\left(\frac{1}{2} + \frac{h}{8\pi e l_\phi^2}\right)]$$

(1)

where $e$ is the elementary charge, $h$ is the Planck constant, $\psi(x)$ is the digamma function, $\varepsilon$ is the parameter correlated to the number of independent channels contributing to the quantum interference, $l_\phi$ is the coherent length which is the maximum length the electron could maintain its phase during the movement. Fig. 1d-f show the HLN fitting results in line with the normalized MC curves at low magnetic field under different temperatures of 2 K, 6 K, and 10 K, respectively, implying that 2D WAL effect in single crystal $Bi_4Br_4$ is indeed evoked by the surface state. The phase coherent lengths $l_\phi$ as a function of temperature derived from the HLN fitting results increase with the decrement of temperature, as plotted in Fig. 1c, indicating that the dephasing is due to inelastic scattering. Additionally, WAL is sensitive to electronic dimensionality, following the formula $l_\phi = T^{-p/2}$, where $p$ describes the dephasing mechanism[36,37]. For 2D system, $p = 1$ and 2 for dephasing mechanism dominated by electron-electron interaction and electron-phonon interaction, respectively[30,38,39]. For 3D system, dephasing mechanism follows $p = 3/2$ and 3 for electron-electron interactions and electron-phonon interactions, respectively[36,37]. The fitting results (black dashed line) in Fig. 1c shows exponential relationship $l_\phi \propto T^{-0.56}$, reconfirming that WAL effect originates from surface state rather than the 3D bulk state.

It is noticeable that both of the WAL effect and WL effect are suppressed with the increment of temperature or magnetic field, as shown in Fig. 1b. This stems from the relationship between the length values of electron mean free path $l$ (the average length that an electron moves between two scattering), magnetic length $l_B = \sqrt{\frac{h}{8\pi eB}}$, spin-orbit scattering length $l_{SO}$ ($l_{SO} \sim 0$ when SOC is extremely strong), and temperature-dependent $l_\phi$. $l_\phi$ is large enough at low temperature, leading to the coherent scattering without the phase change of the electron at low magnetic field due to the fact that $l_B \gg l_\phi \gg l_{SO} \gg l$. The coherent scattering could be eliminated in the condition that $l_\phi$ is less than $l_{SO}$ with increasing temperature, as shown in Fig. 1g, where the phase of the electron changes after two self-crossing scattering loops. When the $l_B$ becomes comparable or smaller to $l_{SO}$ with increasing magnetic field, the Lorentz deflection of charge carriers become dominant and results in the unified behavior of normalized MC[40].

**Facet dependent WL and WAL effect.** Although the observed WAL effect is caused by the 2D orbit motion of charge carriers depending on the magnetic field normal to the *ab* plane, the roles of two surface states in WAL effect still cannot be identified distinguished due to the fact that the normal directions of both the (001) surface and (100) surface have the component of normal direction of *ab* plane. In order to reveal the origin of the WAL effect, we perform the MC measurements at different tilted angles $\beta$, where $\beta$ is the angle between the applied magnetic field and normal direction of *ab* plane, as shown in the inset of Fig. 2a. In this case, the direction of magnetic field is always perpendicular to the current. After subtracting the bulk state contribution using the method stated above, the WAL effect could be clearly identified in all MC curves in the form of sharp cusp with different tilted angles $\beta$, as displayed in Fig. 2a and c. The MC curves as a function of $B$ component normal to (100) surface ($\cos(\beta - 73°)$) and (001) surface ($\cos\beta$) at 2 K are displayed in Fig. 2a and c, respectively, where the special 73 ° is determined by monoclinic structure of Bi$_4$Br$_4$ with the angle 73° between crystal axis *b* and *c*. Fig. 2b shows the enlarged view of the red square in Fig. 2a, where all MC curves under different $\beta$ angles as a function of normal $B$ component to (100) surface are in accord with each other as well as the HLN fitting results at low magnetic field. On the contrary, the discrete behavior of MC curves as a function of normal $B$ component to (001) surface has been identified in Fig. 2d. These results provide solid evidence that the origin of the WAL effect is the 2D (100) surface state.

Compared to the presence of WAL effect at different tilted angles $\beta$, the WL effect, however, is strongly suppressed with the increment of $\beta$ up to 90 °, *i.e.*, magnetic field paralleling to $a$ axis. When the magnetic field is parallel to $a$ axis, the contribution of (001) surface state on MC is negligible since the orbit motion of charge carriers is perpendicular to (001) surface. Therefore, the WL effect shows a strong correlation with the (001) surface state. In order to further confirm the derivation of WL in Bi$_4$Br$_4$, we measure the MC under different angles $\gamma$ between magnetic field and current by confining the direction of magnetic field in $ab$ plane, as shown in the inset of Fig. 3a. There is no normal $B$ component to (001) plane, eliminating the conductivity contribution from (001) surface state. Fig. 2a and c display the MC curves as a function of $B$ component along $a$ axis at 2 K and 10 K, respectively, where the ascending behavior with the increment of normalized $B$ disappears. The absence of WL effect confirms the deduction that the WL effect derives from the (001) surface. Additionally, the WAL phenomenon could be clearly identified at low magnetic field by the HLN fittings shown in Fig. 3b and d at 2 K and 10 K, respectively. Since the normal directions of (001) surface has the constant component of $B\cos\gamma$, the overlapped MC curves in Fig. 3b and d reconfirm the origination of WAL effect from 2D (100) surface state.

For the electron coherent backscattering mostly behaves as constructive interference in quantum diffusion, *i.e.*, WL, in the system with negligible SOC and without the introduction of pseudospin to reserve spin rotational symmetry[30,31,41]. Bi$_4$Br$_4$ possesses strong SOC strength due to the heavy elementary Bi, where the WL effect is expected to be absent. However, the suppressed WL could exist and determined by the energy gap and Berry phase in systems with strong SOC[42,43]. The Hamiltonian including electron spin information can be described by the massive Dirac equation[44]:

$$H = \hbar v(\sigma_x k_y - \sigma_y k_x) + m\sigma_z$$

(2)

where $\hbar$ is reduced Planck constant, $v$ is effective velocity, $\sigma_i$ and $k_i$ denote Pauli matrix and wave vector in *i* direction, *m* is effective mass. The geometric phase (Berry phase) $\varphi$ following the equation[45]:

$$\varphi = \pi\left(1 - \frac{m}{\sqrt{m^2+(v\hbar k_F)^2}}\right) = \pi\left(1 - \frac{\Delta}{2E_F}\right)$$

(3)

where $k_F$ is Fermi wave vector, $\Delta$ is energy gap, $E_F$ is energy difference between Fermi level and middle of energy gap. The linear energy-momentum dispersion without band gap between conduction band and valence band breaks spin rotational symmetry and corresponds to nontrivial Berry phase $\varphi = \pi$ in the condition of *m* = 0. On the contrary, if m≠0, there is an energy gap shown up, violating spin rotational symmetry and bringing Berry phase $\varphi \neq \pi$. There are three kinds of 2D symmetric systems giving rise to different quantum diffusion corrections: orthogonal (elastic) symmetry with both TRS and spin rotational symmetry, unitary (magnetic) symmetry with the absence of TRS, and symplectic (spin-orbit) symmetry with TRS and without spin rotational symmetry[31,46]. The symmetry of Bi$_4$Br$_4$ is attributed to symplectic type because of its large SOC strength. The impurities scattering characteristics of 2D quantum diffusive transport of Dirac fermions in symplectic symmetry are reflected by WAL or suppressed WL depending on Berry phase $\varphi$, where the WAL effect is correlated to the massless Dirac fermion with $\varphi = \pi$, and the WL effect is induced by massive Dirac state with the finite energy gap $\Delta$ and $\varphi \neq \pi$. As stated above, the WAL/ WL effects stem from the (100)/(001) surface state, implying that (100) surface and (001) surface of Bi$_4$Br$_4$ possess the massless Dirac cone structure and gapped band structure without spin rational symmetry, respectively. The schematic diagram of band structures, quantum diffusion corrections, and carriers Berry phases of (100) surface and (001) surface is displayed in Fig. 4a, where a facet dependent topological phase transition is observed in single crystal Bi$_4$Br$_4$.

**The electronic structure of (001) surface and (100) surface.** Angle-resolved photoemission spectroscopy (ARPES) is a powerful tool to directly detect the surface band structure. In order to confirm the deduction of facet dependent topological phase transition by quantum diffusive transport measurements, we perform the ARPES measurements on both of the cleaved (100) and (001) surfaces due to the weak according to the bulk and projected surface Brillouin zones

(BZ) of $Bi_4Br_4$ in Fig. 4b. The high-quality ARPES spectra of (001) and (100) surfaces are shown in Fig. 4c and d, respectively, on the basis of the weak van der Waals interlayer interaction of both facets. The ARPES intensities of constant energy counters (CEC) at different binding energies of (001) surface are plotted in Fig. 4c. One oval-like electron pocket is observed around $\bar{M}$ point of BZ, which firstly shrinks and then frows in area with the increment of binding energy. Another pocket emerges at the center of BZ ($\bar{\Gamma}$ point) with the binding energy larger than 0.6 eV, and then overlaps and hybridizes with the signal from $\bar{M}$ point to form a highly anisotropic band dispersion at deeper energy levels. This high anisotropy stems from the weak van der Waals intralayer interaction along the direction perpendicular to the chain. The energy-momentum dispersions of (001) surface at two time-reversal invariant momenta (TRIM), $\bar{M}$ point and $\bar{\Gamma}$ point of BZ, along the chain direction ($k_y$) are displayed in the right panel of Fig. 4c, where the low-energy bands only distribute near the $\bar{M}$ point of BZ with a bandgap around 0.23 eV between upper band and lower band. The previous research demonstrates that the gap observed in $\bar{M}$ point is quantum spin Hall gap, resulting from the SOC-induced band inversion[13]. Furthermore, the Berry phase $\varphi$ of electron at (001) surface is calculated to be around $0.47\pi$ according to the Equation 3. The SOC related band gap and $\varphi$ value of (001) surface state is well in line with the requirement of the suppressed WL effect in the symplectic symmetry with partial violation of spin rotational symmetry.

The ARPES intensities of CEC at different binding energies of (100) surface are displayed in the left panel of Fig. 4d, where the quasi-1D dispersion along $k_z$ direction is identified at all CEC. This quasi-1D intensity indicates the weaker interlayer coupling along $c$ axis compared to that of the $a$ axis with the anisotropic oval band structure, which is consistent with the calculated binding energies between 1D chains along $a$ axis (~70 meV/atom) and $c$ axis (~87 meV/atom)[11]. When the energy near Fermi surface, two parallel lines-like intensities are observed in CEC. The two lines overlap with each other and then split again with the increment of binding energies. In order to further uncover the band structure of (100) surface, the energy-momentum cuts near $\bar{Z}$ and $\bar{\Gamma}$ point of BZ are displayed in the right panel of Fig. 4d. Interestingly, the linear energy-momentum dispersion, *i.e.*, massless Dirac cone, is unambiguously observed at both of the two TRIM points. The massless Dirac cone resides in the position of bulk gap, which is correlated to the surface states of the topological insulators.

Due to the highly anisotropic shape of Dirac cone, the carriers from the (100) surface state follow two spin-momentum locked channels and yield the Berry phase $\varphi = \pi$, promising the observation of WAL effect that requires spin rotational breaking. Our ARPES spectra coincide with quantum diffusive transport results and provide the solid evidence for the facet dependent topological phase transition in $Bi_4Br_4$.

**Discussion**

It should be noted that the previous research implies that the $\alpha'$-$Bi_4Br_4$ is a high-order topological insulator (HOTI) stemming from the shift of inversion center in the unit cell by the laser-based ARPES measurements[20]. The 1D helical hinge states are expected to emerge between (001) surface and (100) surface in 3D HOTI due to the second-order bulk-boundary correspondence[21,47]. This second-order helical hinge states could also evoke the WAL phenomenon due to the linear energy momentum dispersion and $\pi$ value of Berry phase of the carriers. Nevertheless, neither our MC results nor ARPES spectra support the claim of the second-order helical hinge states of a HOTI $Bi_4Br_4$. The MC measurements illustrate that the WAL phenomenon is ideally arranged along the (100) surface state, while the 1D hinge states are along the chain direction (*b* axis). Such 1D hinge states provide the same contribution to either (001) surface or (100) surface, prohibiting the WAL effect aliened merely on (100) surface. Moreover, the 1D hinge states are expected to be detected on both (001) surface and (100) surface with almost the same intensity due to its isotropic dispersion along *a* axis and *c* axis. The intensity of massless Dirac cone is strong in Fig. 4d ((100) surface), while no related trace could be observed in Fig. 4c ((001) surface). More work, such as the comparative study of scanning tunneling microscopy and spectroscopy on the local electronic structure between (001) surface and (100) surface of $Bi_4Br_4$, are required to clarify this controversy.

In conclusion, we successfully identified the WL effect and WAL effect in $\alpha'$-$Bi_4Br_4$ and attributed their origins to the (001) surface state and (100) surface state, respectively. The different Berry phases of carriers contributing to WL and WAL phenomena indicates the facet dependent topological phase transition in this system, which has been confirmed by the ARPES spectra of both cleaved (001) surface and (100) surface. Our work provides the connection between surface-selected quantum diffusive phenomenon and topological features, shedding light on the exploration of topological mechanism of this novel quasi-1D material.

**Methods**

***α'*-Bi$_4$Br$_4$ growth.** Single crystal *α'*-Bi$_4$Br$_4$ were grown by solid-state reaction method, in which highly pure Bi and BiBr$_3$ powders with equal molar ratio were mixed under Ar atmosphere in a glove box and then sealed in the quartz tube under vacuum. The sealed raw materials were placed in a two-heating-zone furnace with a temperature gradient from 558 K to 461 K for a duration of 72 h. After cooling down, the single crystal Bi$_4$Br$_4$ nucleates at the hot side of the quartz tube.

**Transport measurements.** Selected long and homogeneous *α'*-Bi$_4$Br$_4$ sample with proper size (length ~ 2 mm, width ~ 0.2 mm, thickness ~ 0.1 mm) was firstly exfoliated by scotch tape. The standard four-probe technique under a magnetic field up to 9 T was performed by Physical Property Measurement System (Quantum Design) with rotational rod accessories.

**ARPES characterization.** Selected thick and homogeneous *α'*-Bi$_4$Br$_4$ sample was sticked by torr seal glue with exposed (001) or (100) surface, and then was transformed into ARPES chamber to make a *in situ* cleavage at 6 K. The ARPES characterizations were performed at the Photoelectron Spectroscopy Station in the Beijing Synchrotron Radiation Facility using a SCIENTA R4000 analyzer. A monochromatized He I light source (21.2 eV) was used for the band dispersion measurements. The total energy resolution was better than 15 meV, and the angular resolution was set to ~0.3°, which gives a momentum resolution of ~0.01 π/a.

## Acknowledgements

The work was supported by the Beijing Natural Science Foundation (Z180007) and the National Natural Science Foundation of China (11874003, 11904015, 12005251, 12004321, 12074021).

## Author contributions

M.Y. and F.Y. carried out the single crystal fabrication. J.Y.Z. performed transport measurements and analysis. M.Y. and C.L. performed the ARPES measurements and analysis. J.O.W. participated in the interpretation of the experimental data. J.C.Z. supervised the project. J.C.Z., W.C.H., and Y.D. designed the experiment and wrote the manuscript. All authors contributed to the scientific discussion and manuscript revisions.

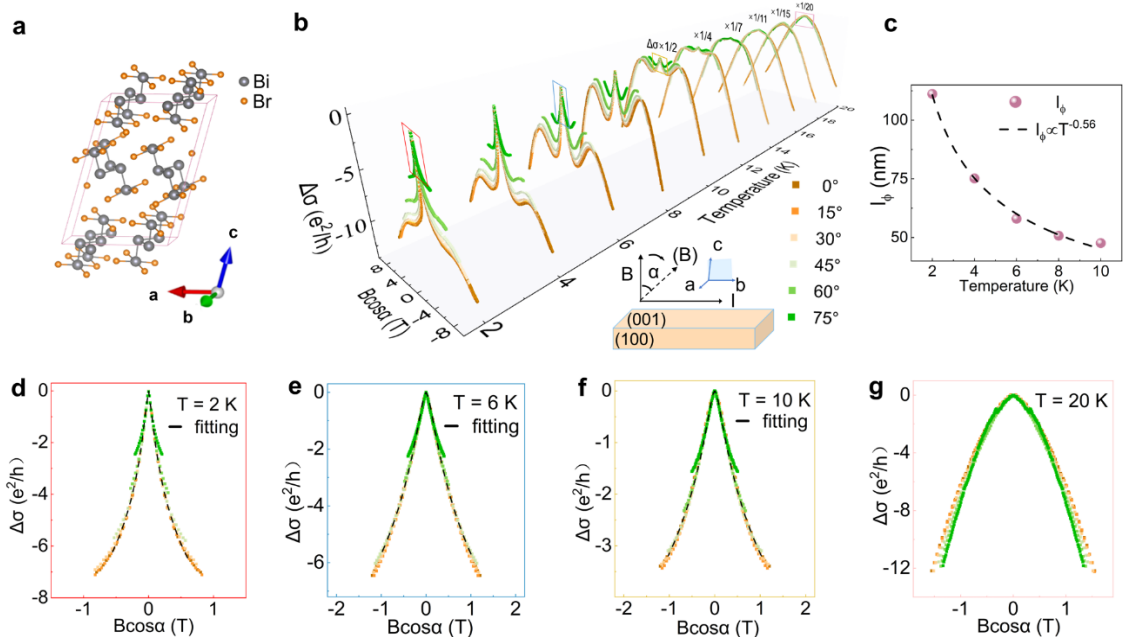

**Fig. 1 Temperature dependent WAL and WL phenomena. a** Schematic diagram of atomic structure of α'-Bi$_4$Br$_4$. **b** Angle-scaled surface MC at temperature from 2 K to 20 K, with the angle α between magnetic field and current varying from 0° to 75°. The MC curves at high temperature have been multiplied by a factor for the eye guidance of comparative study between different curves. The inset of **b** shows the experimental setup of rotation angle α, where the orange cuboid represents the single crystal Bi$_4$Br$_4$, black arrows stand for the directions of magnetic field and current, and blue arrows indicate the crystal axes. **c** Temperature and coherent length relationship shows exponential behavior $l_\phi \propto T^{-0.56}$. **d-f** Enlarged plots with the HLN fitting (black dashed line) from the red, blue and yellow squares in **b**. **g** Enlarged plots from the pink square in **b** measured at 20 K.

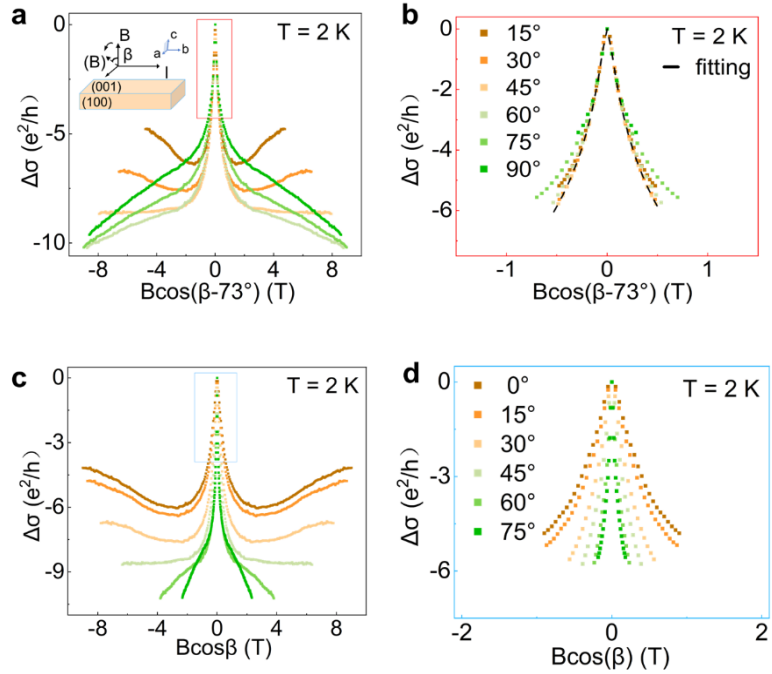

**Fig. 2 (100) surface dependent WAL effect. a** Angle-scaled surface MC measured at different angle *β* between magnetic field and the normal direction of *ab* plane at 2 K. The inset shows the experimental setup of rotation angle *β*. **b** Enlarged plot with HLN fitting from the red square in **a**. **c** MC with *x* axis scaled by normal component of magnetic field towards (001) surface. **d** Enlarged plot from the blue square in **c**.

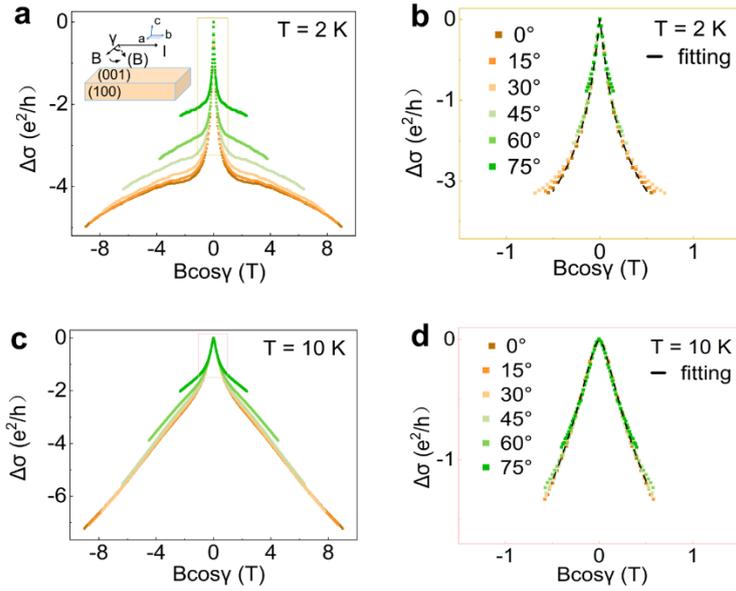

**Fig. 3 (001) surface dependent suppressed WL effect. a** Angle-scaled surface MC measured at 2 K under different angle $\gamma$ between current and the magnetic field restricted in the *ab* plane. The *x* axis is scaled to identify the component of magnetic field along *a* axis. The insetshows the experimental setup of rotation angle $\gamma$. **b** Enlarged plot with HLN fitting from the yellow square in **a**. **c** Angle-scaled surface MC measured at 10 K. **d** Enlarged plot with HLN fitting from the pink square in **c**.

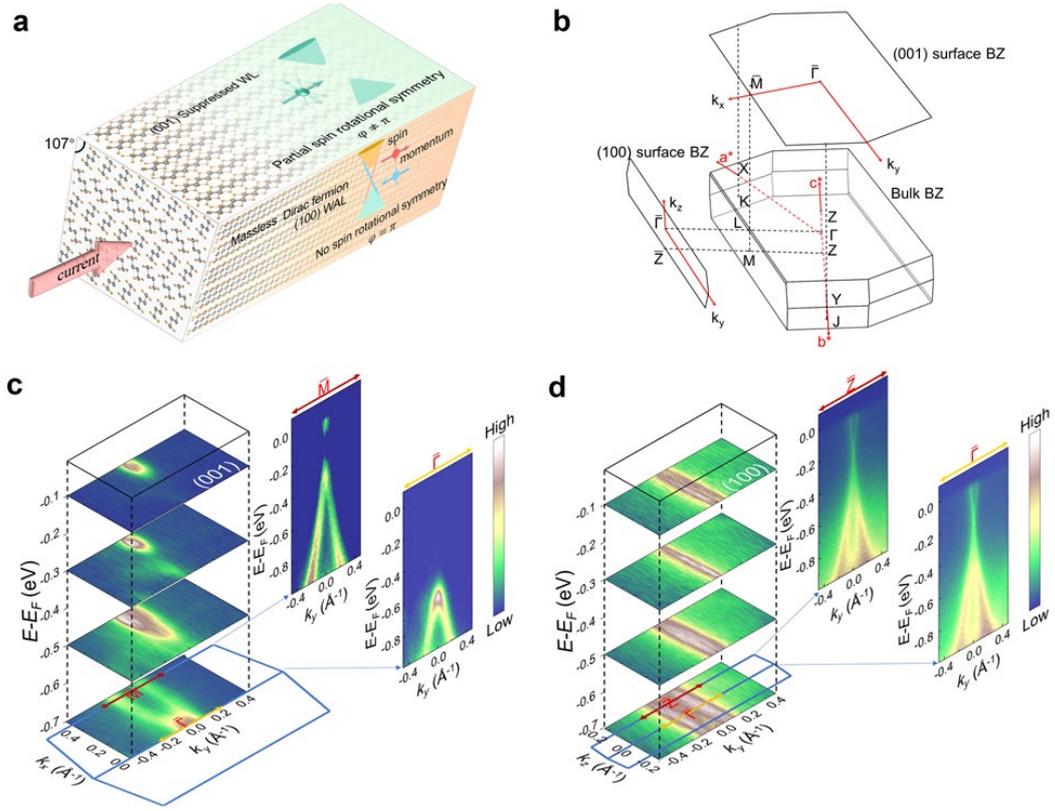

**Fig. 4 ARPES spectra of (001) and (100) surfaces of Bi$_4$Br$_4$. a** Schematic illustration of crystal structure and the facet dependent topological phases, where (001) surface shows suppressed WL phenomenon due to the finite gap with Berry phase $\varphi \neq \pi$, and (100) surface shows WAL phenomenon because of the massless Dirac fermion with Berry phase $\varphi = \pi$. **b** Primitive BZ and TRIM points of bulk Bi$_4$Br$_4$. TRIMs are projected onto the (001) surface of BZ and the (100) surface BZ. **c** Constant energy contour (CEC) of (001) surface at $E$-$E_F$ = -0.1, -0.3, -0.5, -0.7 eV (left part), and band dispersion at TRIMs $\Gamma$ and $M$ along $k_y$ direction measured at $k_x = 0$ and $k_x = 0.43$ (Å$^{-1}$) (right part). **d** CEC of (100) surface at $E$-$E_F$ = -0.1, -0.3, -0.5, -0.7 eV (left part), and band dispersion at TRIMs $\Gamma$ and $Z$ along $k_y$ direction measured at $k_z = 0$ and $k_z = 0.16$ (Å$^{-1}$) (right part).